\documentclass[useAMS,usenatbib,aas_macros]{mn2e}

\usepackage{epsfig}
\usepackage{amssymb}
\usepackage{natbib}
\usepackage{aas_macros}

\begin{document}


\def\fopt{$f_{\rm opt}$}
\def\mfopt{f_{\rm opt}}
\def\mbh{$M_{\rm BH}$}
\def\mmbh{M_{\rm BH}}
\def\sis{$\sigma$}
\def\m{$M$}
\def\mmin{$M_{\rm min}$}
\def\sarc{$^{\prime\prime}\!\!.$}
\def\hmsun{$\, h^{-1}\, {\rm M_{\odot}}$}
\def\arcsec{$^{\prime\prime}$}
\def\lam{$\lambda$}
\def\epsi{$\epsilon$}
\def\arcmin{$^{\prime}$}
\def\degr{$^{\circ}$}
\def\wpr{$W_p(R_p)$}
\def\seco{$^{\rm s}\!\!.$}
\def\ls{\lower 2pt \hbox{$\;\scriptscriptstyle \buildrel<\over\sim\;$}}
\def\gs{\lower 2pt \hbox{$\;\scriptscriptstyle \buildrel>\over\sim\;$}}
\def\ergsec{\,{\rm erg}\,{\rm s}^{-1}}

\title[Black hole duty cycles and clustering]{Constraints on black hole duty cycles and the black hole-halo relation from SDSS quasar clustering}

\author[Shankar et al.]
{Francesco Shankar$^{1}$\thanks{E-mail:$\;$shankar@mpa-garching.mpg.de},
David H. Weinberg$^{2,3}$, and Yue Shen$^{4,5}$\\
$^1$ Max-Planck-Instit\"{u}t f\"{u}r Astrophysik,
Karl-Schwarzschild-Str. 1, D-85748, Garching, Germany\\
$^2$ Astronomy Department and Center for Cosmology and
Astro-Particle Physics, Ohio State University,
    Columbus, OH-43210, U.S.A.\\
$^3$ Institute for Advanced Study, Princeton, NJ, U.S.A.\\
$^4$ Princeton University Observatory, Princeton, NJ, U.S.A.\\
$^5$ Harvard-Smithsonian Center for Astrophysics, Cambridge, MA, U.S.A.
}

\date{}
\pagerange{\pageref{firstpage}--\pageref{lastpage}} \pubyear{2010}
\maketitle


\begin{abstract}
We use Shen et al.'s (2009) measurements of luminosity-dependent clustering
in the SDSS Data Release 5 Quasar Catalog, at redshifts $0.4 \leq z \leq 2.5$,
to constrain the relation between quasar luminosity and host halo mass and to
infer the duty cycle \fopt, the fraction of black holes that shine as optically
luminous quasars at a given time.  We assume a monotonic mean relation between
quasar luminosity and host halo mass, with log-normal scatter $\Sigma$. For
specified \fopt\ and $\Sigma$, matching the observed quasar space density
determines the normalization of the luminosity-halo mass relation, from which
we predict the clustering bias. The data show no change of bias between
the faint and bright halves of the quasar sample but a modest increase in bias
for the brightest 10\%. At the mean redshift $z=1.45$ of the sample,
the data can be well described either by models with small intrinsic scatter
($\Sigma=0.1$ dex) and a duty cycle \fopt$=6\times 10^{-4}$ or by models
with much larger duty cycles and larger values of the scatter. ``Continuity
equation'' models of the black hole mass population imply $\mfopt \geq 2\times
10^{-3}$ in this range of masses and redshifts, and the combination of this
constraint with the clustering measurements implies scatter $\Sigma \geq 0.4$
dex. These findings contrast with those inferred from the much stronger
clustering of high-luminosity quasars at $z\approx 4$, which require minimal
scatter between luminosity and halo mass and duty cycles close to one.
\end{abstract}


\section{Introduction}
\label{sec|intro}

The strong correlations between the masses of central black holes
(BHs) and the luminosities, dynamical masses, and velocity
dispersions $\sigma$ of their host galaxies
imply that the growth processes of BHs and their hosts are
intimately linked
(e.g., Magorrian et al.
1998; Ferrarese \& Merritt 2000; Gebhardt et al.\ 2000;
Ferrarese 2002; Ferrarese \& Ford
2005; Graham 2007; Tundo et al. 2007; Shankar et al. 2009b).
However, constraining the cosmological
evolution of BHs remains a challenge.  Although a variety of
theoretical models
may roughly match observations, the underlying
physical assumptions on BH growth can vary drastically from one model to
another
(e.g., So{\l}tan 1982; Silk \& Rees 1998;
Salucci et al. 1999; Cavaliere \&
Vittorini 2000; Kauffmann \& Haehnelt 2000; Yu \& Tremaine 2002; Steed \& Weinberg 2003;
Wyithe \& Loeb 2003; Granato et al. 2004, 2006; Marconi et
al. 2004; Merloni et al. 2004; Yu \& Lu 2004; Miralda-Escud\`{e}
\& Kollmeier 2005; Murray et al. 2005; Cattaneo et al. 2006;
Croton et al. 2006; Hopkins et al. 2006; Lapi et al. 2006; Shankar
et al. 2004, 2006, 2009a; Malbon et al. 2007; Monaco et al. 2007;
Croton 2009; Cook et al. 2009).
Quasar clustering provides additional, independent constraints on
the BH population, helping to discriminate among
otherwise viable
models. As outlined by Martini \& Weinberg (2001)
and Haiman \& Hui (2001; see also Wyithe \& Loeb 2005;
Lidz et al. 2006; Hopkins et al.
2007a; White et al. 2008; Shen et al. 2009a,b; Shankar et al. 2009c; Wyithe \& Loeb 2009; Bonoli et al. 2009),
the clustering is an
indirect measure of the masses, and therefore number densities, of the
halos hosting the quasars. In turn, the ratio between the quasar
luminosity function and the halo mass function provides information
on the duty cycle, i.e., the fraction of halos that host active
quasars at a given time.
In general terms, stronger clustering implies that quasars reside in
rarer, more massive hosts, and matching the observed quasar space
density then requires a higher duty cycle.

In this paper, we model Shen et al.'s (2009a; S09 hereafter)
recent measurements of luminosity-dependent quasar clustering
derived from the quasar redshift survey (Schneider et al.\ 2007)
of the Sloan Digital Sky Survey (SDSS; York et al. 2000) Data Release 5
(DR5; Adelman-McCarthy et al.\ 2007).
Ross et al.\ (2009) also analyze the clustering of this quasar survey,
concentrating on redshift evolution, but here we focus on the S09
results because they isolate the luminosity dependence of clustering.
Our aim is to answer basic questions about the evolution of
the AGN and supermassive BH population at $z \leq 2.5$.
Does the duty cycle depend on
quasar luminosity and/or redshift? What is the
underlying relation between quasar luminosity
and halo mass? Does it have scatter?
More generally, what combinations of duty cycle and scatter are
allowed by the measurements?

Throughout
the paper we adopt $\Omega_m=0.26$, $\Omega_\Lambda=0.74$,
$h\equiv H_0/100\, {\rm km\, s^{-1}\, Mpc^{-1}}=0.7$,
$\Omega_b=0.0435$, $n_s=0.95$, $\sigma_8=0.78$, and the transfer
function of Eisenstein \& Hu (1999; with zero neutrino
contribution), which matches the cosmology used by S09.

\begin{figure*}
\includegraphics[width=17.5truecm]{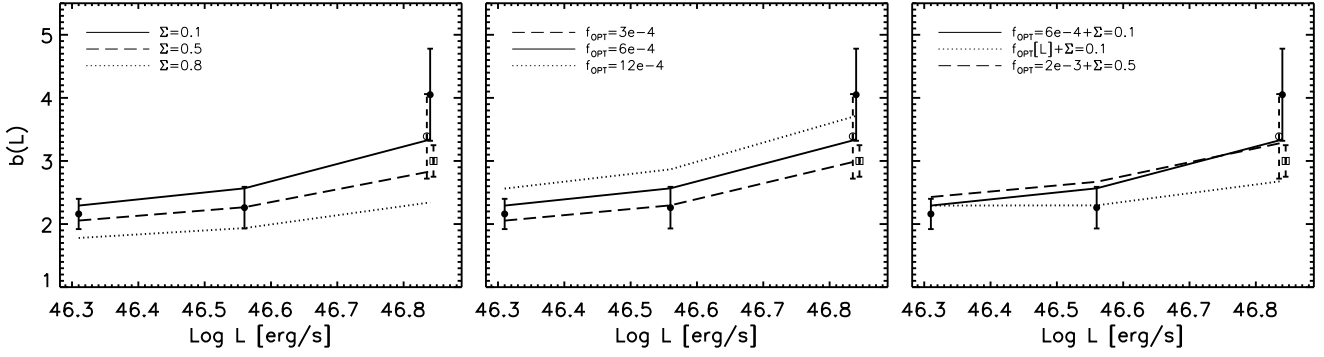}
\caption{Bias as a function of bolometric
luminosity. In all panels, the \emph{solid circles} are the mean
bias measured by S09 from the quasar auto-correlation function
for the faint, bright, and
brightest subsamples in their analysis. The \emph{open circle} with
\emph{dashed} error bars is the bias measured for the brightest
subsample including in the fits the bins with negative correlation
function. The \emph{open square} is the bias computed
from the cross-correlation of the most luminous sources
with the rest of the sample. The data are compared with
predictions of several models for the mean bias at $z=1.45$, the
average redshift of the S09 sample. \emph{Left panel}:
Comparison among models with the same value for the duty cycle
\fopt$=6\times 10^{-4}$
and different values of the intrinsic Gaussian scatter
$\Sigma$ (in dex) in the quasar luminosity-host halo relation, as
labeled. \emph{Central panel}: Comparison among models with the same
scatter $\Sigma=0.1$ dex but different values of the duty cycle \fopt,
as labeled. \emph{Right panel}: Comparison among three different
models: one with
constant \fopt$=6\times 10^{-4}$ at all luminosities
and scatter $\Sigma=0.1$
(\emph{solid} line); another with equal scatter but with a decreasing duty cycle
\fopt$=6\times 10^{-4}$ at $\log (L/{\rm erg\, s^{-1}})=46.31$ and \fopt$/2$
and \fopt$/4$ at $\log (L/{\rm erg\, s^{-1}})=46.56$ and 46.84,
respectively (\emph{dotted} line);
and finally a model with \fopt$=2\times 10^{-3}$ and
$\Sigma=0.5$ (\emph{long-dashed} line). \label{fig|bLz}}
\end{figure*}

\section{Data}
\label{sec|data}

\begin{figure*}
\includegraphics[width=17.5truecm]{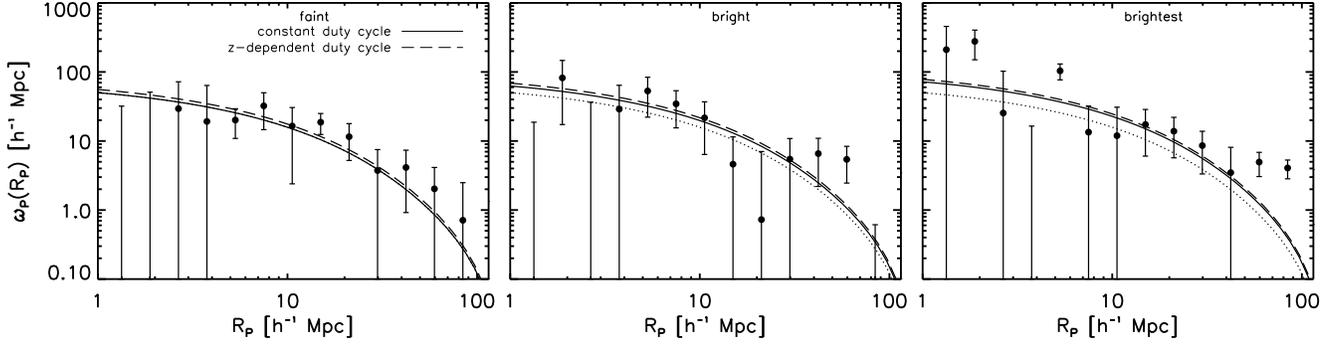}
\caption{The projected correlation
functions from S09 (\emph{solid points} with error bars) are
compared with the predictions of the low-scatter model from
Figure~\ref{fig|bLz}.
The \emph{left}, \emph{middle}, and \emph{right} panels refer to
the faint, bright, and brightest subsamples (with the
cross-correlation shown for the brightest sample). The \emph{solid} lines refer to a model with constant
duty cycle \fopt$=6\times 10^{-4}$ at all luminosities and
redshifts, while the \emph{long-dashed} lines refer to a model
with a redshift-dependent duty cycle \fopt$=6\times
10^{-4}\times(1+z/2.45)^8$.
The predictions of the constant
duty cycle model for the faint sample are also shown as
\emph{dotted} lines in the last two panels for comparison.
\label{fig|WpYue}}
\end{figure*}

The sample used by S09 is a homogeneous subset of a
catalog of 77,429 spectroscopically identified quasars brighter
than $M_i=-22$, in the redshift range $0.1\lesssim z \lesssim
5.0$. Shen et al. (2007) computed the correlation function of the
high-redshift quasars at $z\ge 2.9$, modeled subsequently by White et
al. (2008) and Shankar et al. (2009c). Here instead we focus on
the correlation function of lower redshift quasars in the range
$0.4\le z \le 2.5$. To
probe the luminosity dependence of the bias, S09 divided the
low-$z$ sample into subsamples containing the fainter half of the
quasars, the brighter half of the quasars, and the brightest 10\%
of the quasars (see their Fig.\ 2). In each luminosity bin, they
computed the quasar correlation function. In particular, S09 estimated
for the full sample a mean clustering
bias of $b=2.16\pm 0.24, 2.26\pm 0.33, 4.05\pm 0.73$ for
the faint, bright, and brightest subsamples, with median luminosity
$\log L_{\rm med}/{\rm erg\, s^{-1}}=46.31, 46.56, 46.84$,
respectively.\footnote{We here use the S09 conversion to
bolometric luminosities $L=10^{[M_i(z=2)-90]/(-2.5)}$, with
$M_i(z=2)=M_i(z=0)-0.596$, the $i$-band, $z=2$ $K$-corrected
magnitude system introduced by Richards et al. (2006).}
We will first compare with their data on the bias by computing
models at the average redshift $z=1.45$ of their sample (Figure~\ref{fig|bLz}).
We will then compute the full correlation functions for the faint, bright, and
brightest subsamples averaged over the full redshift distribution
of the sample, and compare them with the S09 measurements
(Figure~\ref{fig|WpYue}).

\section{Method}
\label{sec|method}

By imposing a cumulative
match between the space densities of
quasars and their host halos, and assuming that only a
fraction \fopt\ of halos of a given mass shine as optical quasars
at a given time, we can estimate the mean host halo mass given the
observed number density of quasars. Formally, this concept reads
as (e.g., White et al. 2008)
\begin{eqnarray}
\int_{x_{\rm min}}^{\infty} n(x,z)dx= \int_{-\infty}^{\infty}dy f_{\rm opt}\Phi(y,z)\times \nonumber\\
\frac{1}{2}{\rm erfc}\left[\ln
\left(\frac{10^{y_{\rm min}(x_{\rm min})}}{10^y}\right)\frac{1}{\sqrt{2}\ln(10)\Sigma}\right] ,
\label{eq|cummatching}
\end{eqnarray}
with $x=M_i(z=2)$ and $y=\log M$.
Here $\Phi(y,z)$ is the comoving number density of halos,
in units of ${\rm Mpc^{-3}}\, {\rm dex}^{-1}$ for
$H_0=70\, {\rm km\, s^{-1}\, Mpc^{-1}}$, which we take from Sheth
\& Tormen (1999), while $n(x,z)$ is the comoving number density
of quasars (in ${\rm Mpc^{-3}}$) with absolute
magnitude in the range $x \rightarrow x+dx$. We take the observed luminosity
function $n(x,z)$ from Richards et al. (2006), corrected to our cosmology.
The quantity \fopt\ in Eq.~(\ref{eq|cummatching}) is the duty
cycle, i.e., the fraction of halos that host quasars shining above
a minimum luminosity $x_{\rm min}=M_{i,{\rm min}}$ at redshift $z$.
Eq.~(\ref{eq|cummatching}) also takes into account
a lognormal scatter with dispersion
$\Sigma$ (in dex) around the mean quasar luminosity-halo mass
relation.\footnote{Note that when comparing with the S09
measurements, Eq.~(\ref{eq|cummatching}) should have an upper limit
on the left-hand side corresponding to the maximum luminosity
$M_{i,{\rm max}}$ considered in the clustering measurement of a
given subsample. However, we have checked that, as long as the
right-hand side also has a similar cut-off in halo masses above the
halo mass corresponding to $M_{i,{\rm max}}$, our results do not
change.}.
This scatter includes both the scatter between BH mass and halo mass
and the scatter between luminosity and BH mass (i.e., in the Eddington
ratio), and our analysis does not distinguish the two contributions.

At each redshift, Eq.~(\ref{eq|cummatching}) defines the minimum
halo mass $y_{\rm min}$ corresponding to the minimum luminosity in the
sample $x_{\rm min}$ (the latter taken from S09). We then compute the mean bias $\bar{b}$ associated to a given subsample at redshift $z$ with \emph{median} luminosity $\langle x \rangle=M_{i, {\rm med}}$ and minimum
luminosity $x_{\rm min}$ as
\begin{equation}
\bar{b}_{\langle x \rangle}(z)=\frac{\int_{0}^{\infty}dy\Phi(y,z)W[y_{\rm min}(x_{\rm min}),y]b(y,z)}{\int_{0}^{\infty}
dy\Phi(y,z)W[y_{\rm min}(x_{\rm min}),y]}\, ,
    \label{eq|beff}
\end{equation}
with
\begin{equation}
W[y_{\rm min}(x_{\rm min}),y]={\rm erfc}\left[\ln
\left(\frac{10^{y_{\rm min}(x_{\rm min})}}{10^y}\right)\frac{1}{\sqrt{2}\ln(10)\Sigma}\right]
\label{eq|defErrc}
\end{equation}
and $b(y,z)$ the halo bias given by Sheth et al. (2001).
We stress here that an upper luminosity limit to the bin, corresponding
to an upper cut in halo mass (see footnote 2), does not
significantly alter the expected bias given by Eq.~(\ref{eq|beff}).

To perform a detailed comparison with the S09
data, for at least some of the models discussed below,
we also compute the quasar auto- and cross-correlation functions
for each of S09's redshift and luminosity bins. The quasar
auto-correlation function is given by
\begin{equation}
\xi(R,z)=D^2(z)\bar{b}_{\langle x \rangle}^2(z)\xi_m(R)\, ,
    \label{eq|xiRz}
\end{equation}
where $D(z)$ is the linear growth factor of perturbations, and
$\xi_m(R)$ is the linear matter correlation function at $z=0$
derived from the power spectrum. To compute the cross-correlation
function to compare with the S09 10\% most luminous quasars we
instead use the relation
\begin{equation}
\xi_{\rm cross}(R,z)=D^2(z)\bar{b}_{\rm brightest} \bar{b}_{\rm faint} \xi_m(R)\, ,
    \label{eq|xiRz}
\end{equation}
where
$\bar{b}_{\rm brightest}$ 
is the bias associated to the most luminous quasars, while
$\bar{b}_{\rm faint}$ 
is the average bias associated to
the faintest luminosity in the sample at the same redshift.

We then convert the auto-correlation function into a projected
correlation function via the relation
\begin{equation}
w_P(R_P,z)=2\int_0^{\infty}dR_z\, \xi \left(R=\sqrt{R_p^2+R_z^2},z
\right)\, .
    \label{eq|wPz}
\end{equation}
Finally, we compute the average projected correlation function by
weighting with the quasar number redshift distribution $N(z)$ and
volumes probed in each bin considered as
\begin{equation}
w_P(R_P)=\frac{\int dz \, (dV/dz)\,  N^2(z)w_P(R_P,z)}{\int dz\,
(dV/dz)\,  N^2(z)}\, .
    \label{eq|wPzAve}
\end{equation}
By comparing the bias and the average projected correlation function
with the data, we can extract useful information on the underlying
duty cycle \fopt\ and scatter $\Sigma$.
We have checked that including subhalos as quasar hosts, with
the methods of Giocoli et al.\ (2008), does not noticeably
alter our predicted quasar bias or correlation function, because
the abundance of massive subhalos is very small compared to
the abundance of halos above the minimum halo masses probed here.

\section{Results}
\label{sec|results}

\subsection{Clustering Constraints on Duty Cycle and Scatter}
\label{subsec|results}


Figure~\ref{fig|bLz} compares the bias factor predicted
by several illustrative models to the values inferred by S09 from the quasar
correlation function for the faint, bright, and brightest
subsamples, shown by the solid circles. At each redshift, these
subsamples contain the fainter half, brighter half, and brightest
10\% of the quasars above the SDSS magnitude threshold. The open
square shows the bias factor inferred from the cross-correlation
of the brightest sample with the remaining quasars. The open
circle with dashed error bars shows the bias measured for the
brightest subsample (derived by S09 directly
from the cross-correlation function using their Eq.~[3]) when
negative correlation function points are included in the bias fit.
As discussed by S09, it is unclear whether the negative points are
purely statistical fluctuations or artifacts of the redshift
variation of quasar selection efficiency, so there is some
ambiguity about whether it is more accurate to retain or omit
these data points.  This question should be resolved by the larger
data sample from the final SDSS data release, which will have
smaller statistical fluctuations. Nevertheless,
we will show below that our main
conclusions hold irrespective of the exact
data set considered.

Lines in Figure~\ref{fig|bLz} show model predictions for a
variety of assumptions.
By applying
Eqs.~(\ref{eq|cummatching}) and (\ref{eq|beff}) we compute the mean
bias as a function of bolometric luminosity at the single redshift
$z=1.45$, the mean redshift of the S09 sample, for different input
duty cycle \fopt\ and scatter $\Sigma$. The solid line in the left
panel shows a reference model consistent with
the data on the bias, defined by a small scatter $\Sigma=0.1$ dex and a
constant duty cycle \fopt$=6\times 10^{-4}$. As expected, increasing
the scatter to, e.g., $\Sigma=0.5, 0.8$, lowers the predicted bias
and flattens the relation between bias and luminosity
(long-dashed and dotted lines),
as it increases the contamination by the much more numerous, less
massive, and less biased halos. The corresponding effective halo
masses $M_{\rm eff}$ for the faint, bright, and brightest subsamples
are computed by solving the equation $b(M_{\rm eff},z)=\langle
b\rangle(z)$, which yields $M_{\rm eff}\sim 2\times 10^{12}$\hmsun,
$\sim 3\times 10^{12}$\hmsun, and $\sim 10^{13}$\hmsun,
respectively, for the reference model $\langle b \rangle$ values.
The central panel of Figure~\ref{fig|bLz}
compares models with the
same scatter $\Sigma=0.1$ dex but different values of the duty cycle
\fopt, as labeled. Increasing \fopt\ implies
mapping the same number of quasars to less numerous halos (cfr.
Eq.~[\ref{eq|cummatching}]), which are more massive and more biased,
thus inducing an overall increase in the average predicted bias.
Just the opposite is true if the duty cycle is decreased.

\begin{figure*}
\includegraphics[width=9.5truecm]{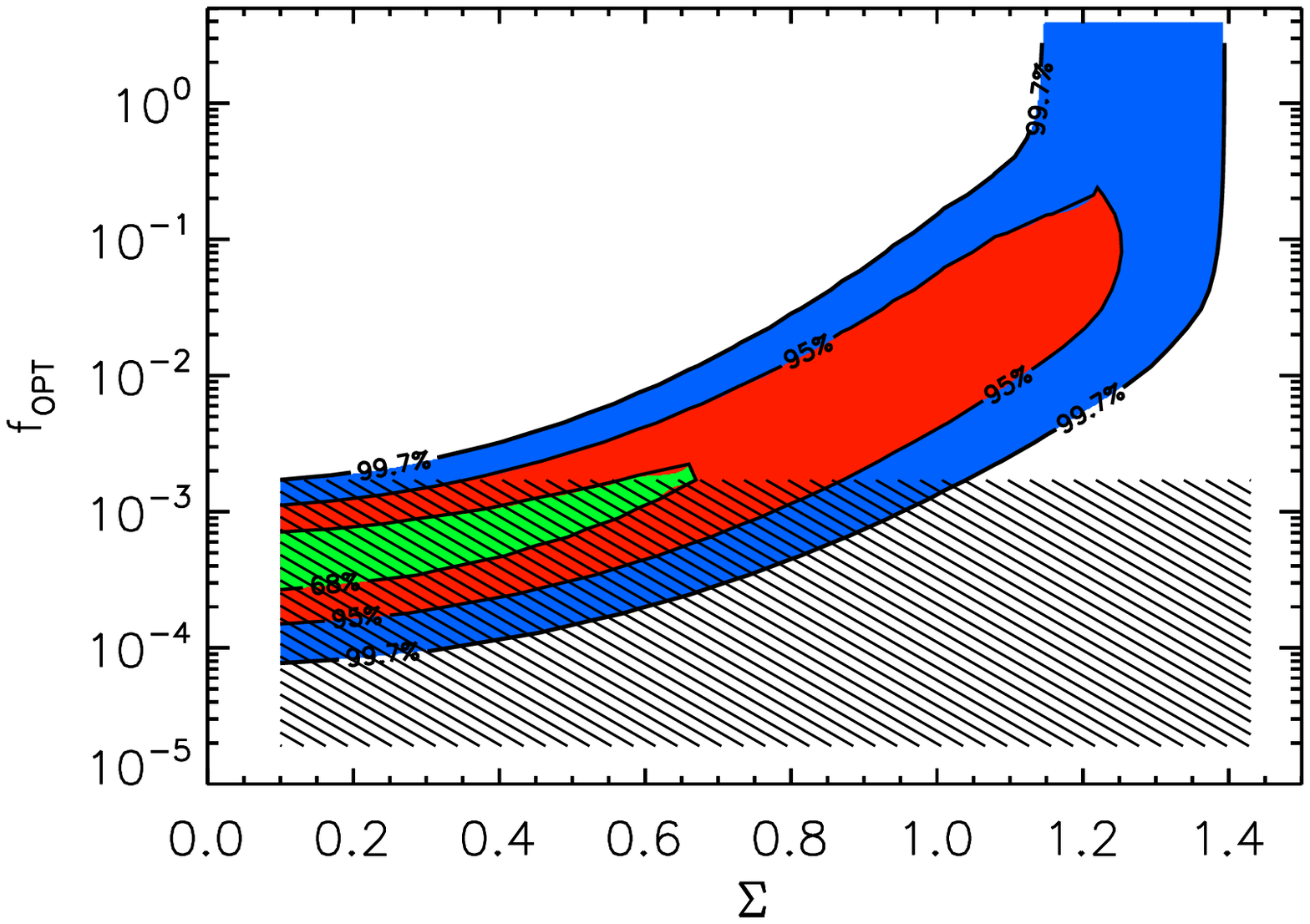}\hspace{-2.2cm}
\includegraphics[width=9.5truecm]{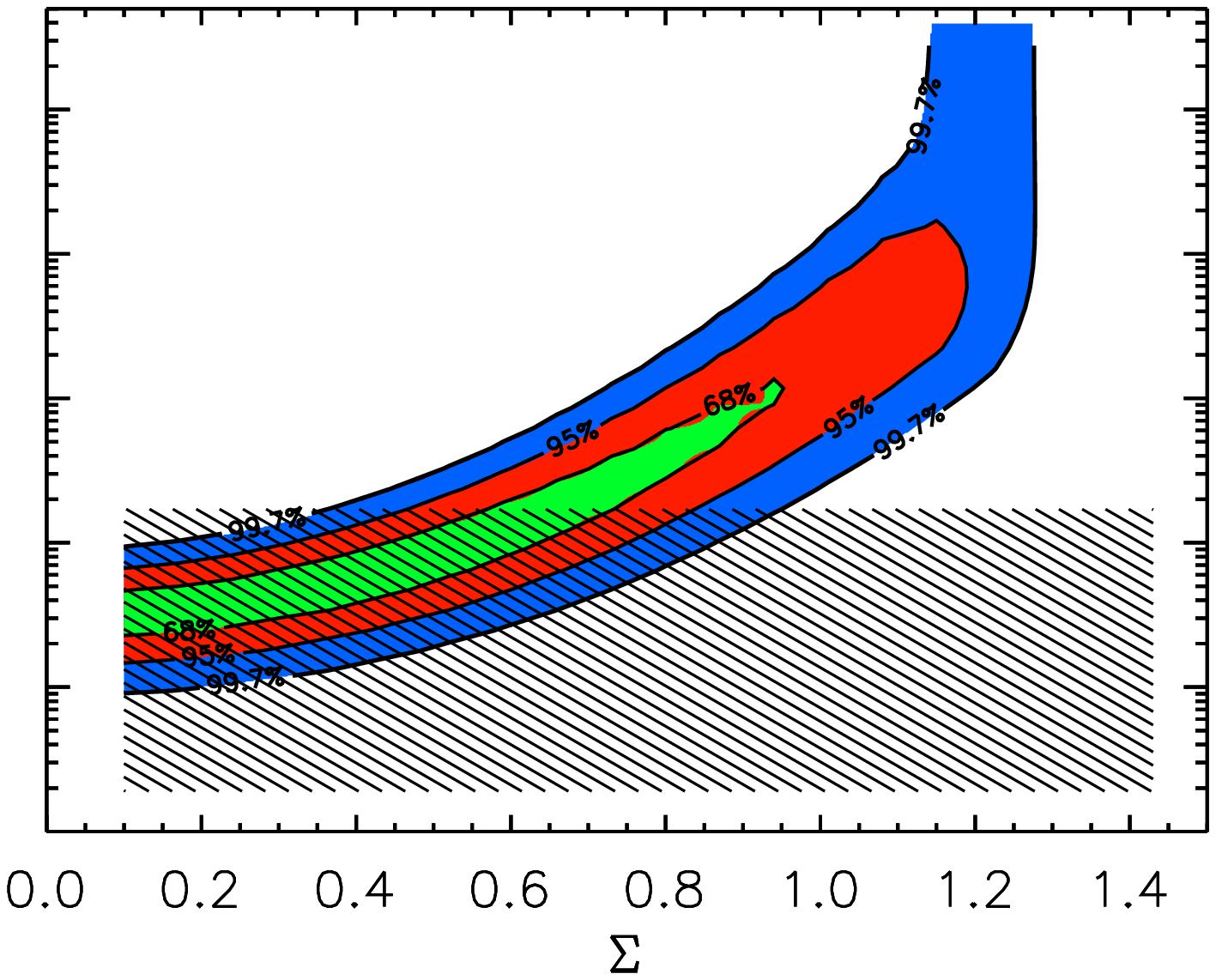}
\caption{Constraints on the optical duty cycle $\mfopt$ and
luminosity-halo mass scatter $\Sigma$ derived from the $b(L)$
data points shown in Fig.~\ref{fig|bLz}.
Contours mark the regions of parameter space with $\chi^2=1.0$,
4.0, and 9.0, corresponding to 68\%, 95\%, and 99.7\% confidence
levels for one degree of freedom (three data points minus two parameters).
Results in the left panel adopt the autocorrelation $b(L)$ estimate
for the highest luminosity bin (rightmost solid circle in
Fig.~\ref{fig|bLz}), while results in the right panel adopt the
cross-correlation estimate (open square in Fig.~\ref{fig|bLz}),
which has a lower central value and smaller error bar.
In both panels, shaded regions indicate the duty cycles
$\mfopt < 2\times 10^{-3}$ that are inconsistent with
the expectations from continuity-equation models of the
black hole population (see \S\ref{subsec|conteq}).
\label{fig|chi2}}
\end{figure*}

The S09 bias measurements are consistent with a constant duty
cycle \fopt$= 6 \times 10^{-4}$ and small scatter
$\Sigma = 0.1$ dex, though this model is $0.5-1\sigma$ high in the
faint and bright bins and $1\sigma$ low (compared to the solid
circle) in the brightest bin.
Allowing an increase in \fopt\ with increasing luminosity would slightly
improve the match to the data. On the other hand, the right
panel of Fig.~\ref{fig|bLz} shows that a model with a significantly
decreasing duty cycle (dotted line), equal to \fopt$=6\times
10^{-4}$ at $\log (L/{\rm erg\, s^{-1}})=46.31$
and to \fopt$/2$ and \fopt$/4$ at
$\log (L/{\rm erg\, s^{-1}})=46.56$ and $\log (L/{\rm erg\,
s^{-1}})=46.84$, respectively,
is inconsistent with the autocorrelation bias for the highest
luminosity bin at the $\sim 2\sigma$ level.
However, if we take the S09 bias measurement that includes negative
data points, or the cross-correlation measurement,
then the discrepancy is only $\sim 1\sigma$. A model
characterized by a high duty cycle and a larger
scatter in the luminosity-halo relation (long-dashed line) is also
consistent with the data at the $\sim 1\sigma$ level.

To make use of the full data sets available,
we
compute the correlation function for a subset of representative
models. The solid circles with error bars in the left, central, and right
panels of Fig.~\ref{fig|WpYue} are the projected correlation
functions \wpr\ estimated by S09 for the faint, bright, and
brightest quasar samples, respectively (the cross-correlation
function is shown for the brightest sample). The solid lines in
each panel refer to the prediction of the reference model
discussed in Fig.~\ref{fig|bLz} defined by a constant
\fopt=$6\times 10^{-4}$, with the correlation function computed
via Eq.~(\ref{eq|wPzAve}) by integrating over
the redshift distribution of the clustering sample. We have
verified that simply computing the correlation function at the
mean redshift of $z=1.45$ produces essentially the same result
(the correlation function in the latter case is systematically
lower by only $\sim 3-4\%$, at fixed \fopt). The reference model
agrees well with the $w_P(R_P)$ data at $R_P \leq 40h^{-1}\,$Mpc.
While the increases in the predicted $w_P(R_P)$ for the
bright and brightest samples are modest, they clearly improve the
fit to the data relative to a luminosity-independent $w_P(R_P)$
(dotted lines).

For the bright and brightest samples, the data at larger scales do
not follow the theoretically predicted shape.
Since this shape is generic to $\Lambda$CDM models with scale-independent
large scale bias, and thus to models that accurately describe
observed galaxy clustering at lower redshifts
(e.g., Reid et al.\ 2009), 
we attribute little significance to this discrepancy at present;
error bars in $w_P(R_P)$ are correlated, and the jackknife method
may underestimate them at large scales.

Other methods (see
\S\ref{subsec|conteq} below)
favour duty cycles that evolve in time. The study
by Shankar et al. (2009a) yields a rapidly evolving duty
cycle that can be approximated as
$f=f(z=1.45)\times((1+z)/2.45)^{8}$ in the range $0.5\lesssim
z\lesssim 2$ (see their Fig. 7c). Applying the latter model to
Eqs.~(\ref{eq|wPzAve}) yields the long-dashed lines in
Fig.~\ref{fig|WpYue}, which is very similar to the reference
model computed at $z=1.45$.

Figure~\ref{fig|bLz} demonstrates a tradeoff between $\mfopt$ and $\Sigma$:
the bias decreases with either decreasing duty cycle or increasing scatter.
The contours in Figure~\ref{fig|chi2} present this tradeoff systematically,
showing regions in the $(\mfopt,\Sigma)$ parameter space that are consistent
with the $b(L)$ data at the 1, 2, and $3\sigma$ confidence levels.
For these contours we assume that $\mfopt$ is independent of $L$ and compute
the predicted bias at $z=1.45$.  The two parameters are not completely
degenerate, as raising $\Sigma$ flattens the $b(L)$ relation in addition
to lowering its amplitude.  If we adopt the autocorrelation estimate
of $b(L)$ for the highest luminosity bin (rightmost solid circle in
Fig.~\ref{fig|bLz}, left panel of Fig.~\ref{fig|chi2}), then values of
$\Sigma > 0.65$ are inconsistent at the $1\sigma$ level because they
predict a $b(L)$ relation that is too flat.  However, the lower $b(L)$
estimated from cross-correlation (open square in Fig.~\ref{fig|bLz},
right panel of Fig.~\ref{fig|chi2}) allows higher $\Sigma$ values,
and the $2\sigma$ constraints are weak in either case.
Future bias measurements with smaller uncertainties could help to
break the $\mfopt$-$\Sigma$ degeneracy, but only to the extent that
they clearly demonstrate a luminosity-dependent clustering trend.
For now, we turn to independent constraints on optical
duty cycles derived from the observed space density of quasars
and models of the underlying black hole population.

\subsection{Additional constraints from the black hole continuity equation}
\label{subsec|conteq}

A widely used method to model the accretion history of the BH
population employs a continuity equation
(Cavaliere et al.\ 1972; Small \& Blandford 1992) to
track the growth of the BH mass function that is implied by the
observed quasar luminosity function.
This approach is reviewed extensively by Shankar et al.\ (2009a;
hereafter SWM), who apply it to a compilation of recent data sets,
and whose results and methodology we adopt here.
The parameters of a model are the radiative efficiency $\epsilon$,
which converts an observed luminosity to a corresponding mass
accretion rate, and the Eddington ratio $\lambda=L/L_{\rm Edd}$,
which determines the mass of the BHs to be associated with a
given observed luminosity.  The method can be generalized
to allow a distribution of $\lambda$ values (Shankar 2009).
For a single $\lambda$ value, the duty cycle is simply
\begin{equation}
f(M_{\rm BH},z)=\frac{\Phi(L,z)}{\Phi_{\rm BH}(M_{\rm BH},z)}\, ,
    \label{eq|P0general}
\end{equation}
where $\Phi(L,z)$ is the quasar luminosity function and
$\Phi(M_{\rm BH},z)$ is the BH mass function at the mass
that corresponds to luminosity $L$,
$M_{\rm BH} = 10^8 \lambda^{-1} (L/10^{46.1} \ergsec) M_\odot$.
For a distribution of $\lambda$, one must take some care
in defining the meaning of the term ``active.''
At redshifts $z>1$, BH mergers are expected to play
a minor role in shaping $\Phi(M_{\rm BH},z)$ relative to
accretion (SWM), and we neglect them here.

\begin{figure*}
\includegraphics[width=9.7truecm]{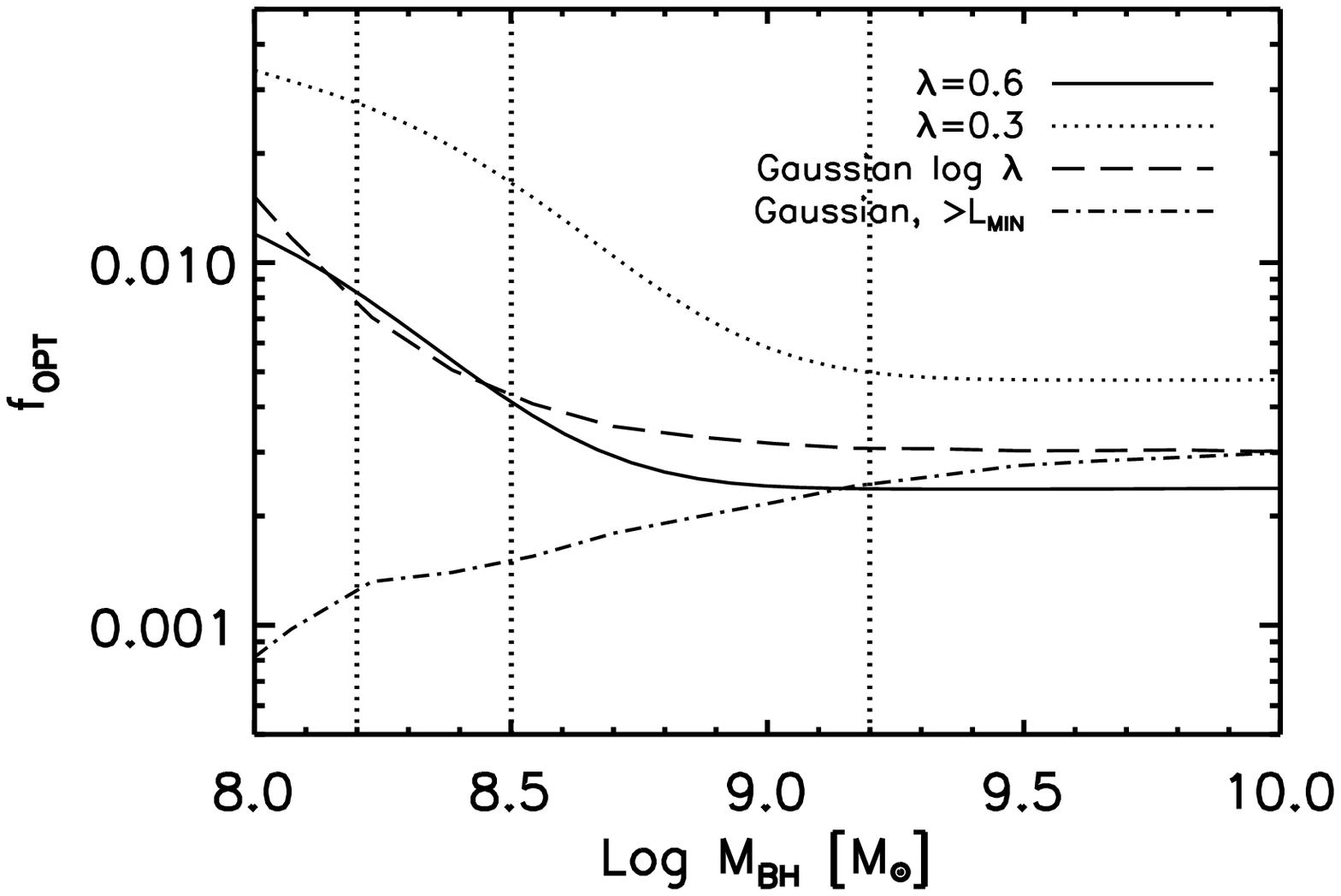}\hspace{-2.7cm}
\includegraphics[width=9.7truecm]{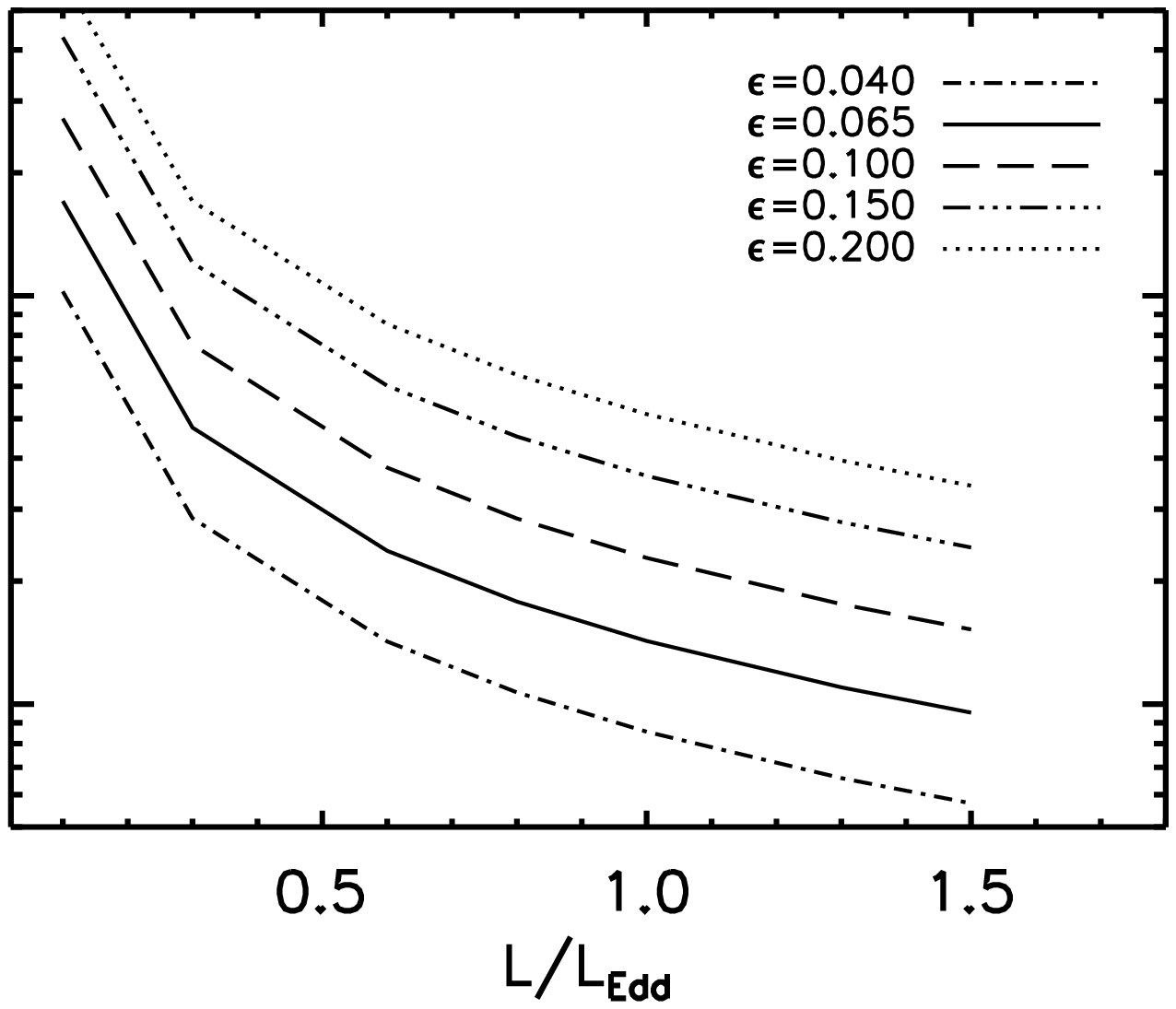}
\caption{Optical duty cycles predicted by continuity-equation
models of the black hole population, as discussed in \S\ref{subsec|conteq},
including a factor of three correction for obscuration, $\mfopt=f/3$.
{\it Left:} Optical duty cycle vs.\ black hole mass at $z=1.45$ for models
with Eddington ratio $\lambda=0.6$ (solid line), $\lambda=0.3$ (dotted line),
and a 0.6-dex Gaussian spread of $\log\lambda$ centered at
$\lambda_{\rm med}=0.3$ (dashed line), all assuming radiative efficiency
$\epsilon=0.065$.
Thick vertical dotted lines show the masses corresponding to the
S09 sample luminosity threshold at $z=1.45$ for
$\lambda=1.0$, 0.5, 0.1 (left to right).
The dot-dashed curve shows the duty cycle in the Gaussian
case with the additional requirement that the Eddington ratio
is high enough to pass this luminosity threshold.
{\it Right:} Optical duty cycle at $\mmbh=10^9M_\odot$ and
$z=1.45$ for models with different choices of $\lambda = L/L_{\rm Edd}$
and $\epsilon$, as indicated.  Within a given model, the
values of $\lambda$ and $\epsilon$ have no scatter and
are held fixed during evolution.
\label{fig|fcont}}
\end{figure*}

The left panel of Figure~\ref{fig|fcont}, analogous to
figure~7c of SWM, shows the optical duty cycle as a function of
black hole mass at $z=1.45$ predicted by several different
continuity equation models.  The model shown by the solid
curve has $\lambda=0.6$ and $\epsilon=0.065$, independent of mass
and redshift, which SWM show yields a good match to observational
estimates of the local black hole mass function.
We convert the total duty cycle to the optical duty cycle
using $\mfopt=f/3$, based on the ratio of the optical luminosity
function for the SDSS quasar sample to the bolometric
luminosity function in SWM.
Above $\mmbh=10^{8.8}M_\odot$, the predicted duty cycle is
$\mfopt=3.6\times 10^{-3}$, while the differing shapes of the
quasar luminosity function and the evolved black hole mass function
imply higher duty cycles at lower masses.
The thick vertical lines mark the masses that correspond to the lower luminosity
limit of the S09 sample at $z=1.45$, $\log\mmbh = 8.2-\log\lambda$.
As discussed by SWM,
including black hole mergers in the mass function evolution or
varying the bolometric luminosity functions or bolometric corrections
within observationally acceptable bounds has minimal impact on
the inferred duty cycles at these redshifts; the largest systematic
uncertainties are associated with the choices of $\lambda$ and $\epsilon$.
The dotted curve shows a model with $\lambda=0.3$, which has
similar shape but higher normalization.  The normalization trend
is easily understood: the integrated quasar
emissivity determines the total mass density of the black hole
population (So\l tan 1982), and assuming lower $\lambda$ shifts this
density to more massive, hence rarer, black holes, which must
have a higher duty cycle to reproduce the luminosity function.
The dashed curve shows a model with a {\it spread} in Eddington
ratios, Gaussian in $\log\lambda$ with 0.6-dex dispersion and
peak at $\lambda_{\rm med}=0.3$, evolved with the techniques
described in Shankar (2009).  Results are intermediate between
the two constant-$\lambda$ models.  However, in this case the
S09 luminosity threshold does not correspond to a sharp mass cut,
so the dot-dashed curve shows $\mfopt$ with the additional
criterion that $\log\lambda > 8.2 - \log\mmbh$, which eliminates
those lower mass black holes whose Eddington ratio would be too
low to enter the S09 sample at $z=1.45$.
This curve is slightly jagged because the calculation uses
a discrete representation of the Gaussian $\log\lambda$
distribution rather than a smooth function (see Shankar 2009).

The right panel of Figure~\ref{fig|fcont} shows the optical
duty cycle at $\mmbh=10^9 M_\odot$ and $z=1.45$ for models
with a range of Eddington ratios and radiative efficiencies.
For lower $\epsilon$, the observed quasar emissivity implies
a higher black hole mass density, hence a higher space density
of black holes at a given mass, and thus a lower duty cycle.
However, observational estimates of the local black hole
mass density imply $\epsilon \ga 0.06$ (see SWM for extensive
discussion), and arguments from accretion disk theory favor
$\epsilon \approx 0.08-0.2$ depending on assumptions about
typical black hole spins (e.g., Berti \& Volonteri 2008, and references therein).
Together with observational estimates favouring Eddington ratios
$\lambda \approx 0.25$ (Kollmeier et al.\ 2006)
or even lower (e.g., Netzer \& Trakhtenbrot 2007) at this luminosity and redshift,
we conclude that continuity-equation models imply optical
duty cycles $\mfopt (z=1.45) > 2\times 10^{-3}$ in the S09
luminosity range.  The robustness of this lower limit depends
on how pessimistically one views the systematic uncertainties
in the local black hole mass density (hence $\epsilon$),
direct black hole mass estimates (hence $\lambda$),
and obscuration fractions (hence $\mfopt/f$), but it is
much easier to find ways to push $\mfopt$ higher than to
push it lower.

Returning to Figure~\ref{fig|chi2}, the shaded bands show the
values of $\mfopt < 2\times 10^{-3}$
excluded by the continuity equation arguments.
Models consistent with this constraint and the 95\% constraint
from the $b(L)$ data have high scatter, $\Sigma > 0.4$ dex.
This conclusion holds regardless of whether we use the autocorrelation
or cross-correlation estimates of the bias in the highest luminosity
bins (left and right panels, respectively).  In the right panel
of Figure~\ref{fig|bLz}, the long-dashed line shows an explicit
example of $b(L)$ for a model with $\mfopt = 2\times 10^{-3}$ and
$\Sigma=0.5$.  The prediction is very similar to that of the
low-scatter model with $\mfopt=6\times 10^{-4}$, though the larger
scatter does produce slightly flatter $b(L)$.

\section{Summary and Implications}
\label{sec|discu}

Studies of quasar clustering have generally failed to find any
significant dependence of clustering strength on quasar
luminosity, at least at $z \leq 2.5$.  The S09 study is one of the
first to separate luminosity dependence from redshift evolution,
and it mostly confirms this basic finding, except for the $\sim
2\sigma$ increase in bias for the brightest 10\% of the quasars
at $z \leq 2.5$.  At first glance,
the absence of luminosity dependence appears to contradict models
like those of Martini \& Weinberg (2001) or Haiman \& Hui (2001),
which assume a monotonic relation between quasar luminosity and
host halo mass and therefore predict a stronger bias for more
luminous quasars. However, Figures~\ref{fig|bLz}
and~\ref{fig|WpYue} show that the S09 results can be reproduced by
a model with constant duty cycle for optical quasar
activity, $\mfopt \approx 6\times 10^{-4}$,
and minimal scatter between luminosity and halo mass.
The weakness of the predicted
luminosity dependence arises because, even with the large size of
the SDSS quasar survey, the dynamic range of luminosity at fixed
redshift is not very large ($\approx 0.5$ dex), and the host halos
at these luminosities and redshifts are not on the extreme,
steeply rising tail of the $b(M)$ relation.  Croton (2009) reaches
a similar conclusion (comparing to other data sets), with a model
that is different in technical implementation from ours but
similar in practice.

However, the S09 bias measurements can also be fit by models
with a higher duty cycle and substantial scatter between
luminosity and halo mass.  The increase in bias for S09's highest
luminosity bin implies an upper limit on scatter, but
this increase is only marginally detected depending on which
method is used to estimate the bias.  Figure~\ref{fig|bLz} shows
an explicit example of an acceptable model with $\mfopt=2\times 10^{-3}$
and log-normal scatter $\Sigma=0.5$ dex, and Figure~\ref{fig|chi2} shows
the regions of the $\mfopt-\Sigma$ parameter space that yield
acceptable agreement with the S09 bias measurements.
As discussed in \S\ref{subsec|conteq}, models of the quasar population
that infer the duty cycle by evolving the BH mass function and
comparing to the quasar luminosity function imply $\mfopt \ga 2\times 10^{-3}$.
Taken together, the clustering constraints and the continuity equation
models imply substantial scatter in the luminosity-halo mass relation,
with $\Sigma \geq 0.4$ dex.

Applying linewidth estimators of BH mass in the AGN and Galaxy Evolution Survey
(AGES), Kollmeier et al.\ (2006) infer a scatter in quasar Eddington ratios
of $\sigma_\lambda \leq 0.3$ dex, though Netzer et al.\ (2007) and
Shen et al.\ (2008) argue for somewhat larger scatter based on other
data sets.  The total scatter between luminosity and halo mass is
a combination (in quadrature) of the scatter in Eddington ratios and
the scatter between halo mass and BH mass.
Physically, many models of quasar activity predict broad Eddington
ratio distributions as a consequence of ``post-peak'' accretion onto
a central BH, after a rapid growth phase in which the BH mass grows
exponentially at a near-Eddington accretion rate
(e.g., Yu \& Lu 2008; Hopkins \& Hernquist 2009; Shen 2009).
Various authors have argued that such prolonged post-peak activity
is the key to reconciling the faint end of the AGN luminosity function
with measurements of quasar bias at low redshift
(e.g., the above papers and Lidz et al.\ 2006; Marulli et al.\ 2008;
Bonoli et al.\ 2009; Shankar et al., in prep.).
We conclude that scatter of $0.4-0.6$ dex in the luminosity-halo mass
relation at these redshifts is plausible on both theoretical
and observational grounds.

Several groups have recently tried to measure, or limit, redshift
evolution of the scaling between BH mass and host galaxy properties.
As several recent papers have pointed out
(e.g., Lauer et al. 2007; Merloni et al. 2010;
Shankar et al. 2009b; Shen \& Kelly 2009),
a large scatter between quasar luminosity and the galaxy scaling property
(such as stellar mass or velocity dispersion $\sigma$) can bias
such measurements.  These biases arise from
a combination of flux-limit effects, rapidly falling stellar
mass (or velocity dispersion) functions of galaxies, and intrinsic
scatter in the scaling relations themselves,
which conspire to cause an apparent rise in the mean
BH mass at fixed galaxy properties with increasing redshift.
Merloni et al. (2010) note
that an increasing scatter with increasing $z$
could be enough to explain the trend
of evolving black hole mass over galaxy mass ratio
measured in their data. Decarli et al. (2010; see also
Bennert et al. 2010 for similar conclusions at lower redshifts)
argue that strong evolution in the black hole mass-galaxy mass relation
is still present even after carefully
accounting for flux-limit effects,
although they did not allow the possibility
of redshift-dependent scatter in the relations (see
also discussion in Shen \& Kelly 2009).
The substantial scatter inferred from our analysis shows that
biases associated with this scatter must be carefully assessed
in studies of the evolution of scaling relations.

A large dispersion between quasar luminosity and host halo mass cannot
be the general rule at all redshifts and luminosities.
In particular, explaining the high clustering amplitude measured for
SDSS quasars at $z \approx 4$ by Shen et al.\ (2007) requires
both minimal scatter and duty cycles close to one
(White et al.\ 2008; Shankar et al.\ 2009c; Bonoli et al.\ 2009).
The quasars in this $z\approx 4$ sample are considerably more luminous than
the lower redshift quasars whose clustering is modeled here,
so in principle the difference in scatter could reflect either
redshift dependence or luminosity dependence.
Fine et al.\ (2008) claim direct empirical evidence for a decrease
of $\Sigma$ with increasing quasar luminosity, based on linewidth
estimates of BH mass, and a decrease of this sort is also found
in numerical simulations of merger-driven quasar activity
(e.g., Hopkins \& Hernquist 2009, and references therein).
Assuming that $\lambda \approx 1$ sets an upper limit on BH luminosity,
decreasing scatter at high luminosity is plausible because the
BH mass function declines rapidly at high masses, so that the
most luminous quasars will almost always be powered by BHs
radiating near the maximum allowed Eddington ratio.
(These arguments address only the scatter in $\lambda$, not
the scatter in BH mass at fixed halo mass.)
We have checked that we can fit the $b(L)$ data in
Figure~\ref{fig|bLz} using models with $\mfopt \approx 10^{-3}$ and
decreasing scatter at high luminosity,
e.g., $\Sigma(L) = 0.6$, 0.3, 0.1 dex for the three bins
of increasing luminosity, or even $\Sigma(L)=0.4$, 0.2, 0.1 dex.
However, the bias in the highest luminosity bin, which is rather
uncertain at present, can significantly constrain such models.

The duty cycles inferred from our analysis at $z\approx 1.45$
are substantially lower than the values $f \approx 0.2$ and
$f\approx 1$ inferred from the Shen et al.\ (2007) measurements
of the clustering of quasars at $z\approx 3$ and $z\approx 4$
(see Shen et al.\ 2007; White et al.\ 2008; Shankar et al.\ 2009c).
This decline in duty cycle at low redshifts is expected from
continuity equation models:
the BH mass function grows in time, but the observed quasar
luminosity function declines at $z < 2$, so a lower duty
cycle is required to reconcile them (see, e.g., figure 7 of SWM).
Our current analysis does not constrain duty cycle evolution
at $z<2$, but strong evolution over this interval is predicted
by the SWM model and is consistent with the S09 correlation
function data (see Figure~\ref{fig|WpYue}).

The measurements in S09 provide significant constraints on
the relation between quasar luminosity and halo mass, though
leaving substantial degeneracy between the duty cycle and the
scatter in this relation.  Reducing
statistical errors and remaining systematic uncertainties, especially
for the brightest luminosities, would tighten these constraints;
in particular,
an unambiguous and precise measurement of luminosity-dependent
bias would place much tighter restrictions on scatter.
The quasar catalog from SDSS DR7 (Adelman-McCarthy et al.\ 2008)
should yield noticeable improvements, with roughly 50\% smaller
error bars and fewer issues with internal boundaries in the survey
region.  Since the SDSS quasar sample has high completeness and
(with DR7) covers most of the high-latitude northern sky,
it will be difficult to go much further with autocorrelation
measurements in the S09 luminosity and redshift range.
Cross-correlation against denser samples of objects ---
fainter AGN or bright galaxies --- could yield higher precision clustering
measurements, perhaps with photometric samples
from surveys such as Pan-STARRS and LSST, but perhaps requiring
spectroscopic samples like those envisioned for ambitious
baryon acoustic oscillation experiments.
The constraints on host halo populations can also be improved
by extending clustering measurements to smaller scales, where quasar
pairs from the same halo contribute, and to fainter luminosities,
such as those probed by the 2dF Quasar Redshift Survey, the SDSS
photometric quasar catalog,
and X-ray surveys (e.g., Hennawi et al. 2006; Myers et al. 2007;
Plionis et al. 2008, Hennawi et al. 2009); for example, Shen et al.\ (2009b)
use small scale measurements to put constraints on the duty
cycle of BHs in satellite galaxies.
Quasar clustering as a cosmological tool has moved from a
prospect (Osmer 1981)
to reality, and the growing precision and dynamic range
of these measurements --- in luminosity, redshift, and lengthscale ---
will teach us about the growth of supermassive
black holes and the mechanisms that transform them from
dormant monsters to brilliant beacons, and back.


\section*{Acknowledgments}
FS acknowledges
the Alexander von Humboldt Foundation for support.
FS and DW also acknowledge support from NASA Grant NNG05GH77G,
and DW acknowledges support of an AMIAS membership at the
Institute for Advanced Study.
We thank Raul Angulo, Luca Graziani, Jeremy Tinker, Jaiyul Yoo,
and Zheng Zheng for interesting and helpful discussions.
We thank the anonymous referee for comments that led to
significant improvements of the paper.

\label{lastpage}

\end{document}